\documentclass[a4paper,10pt,twoside]{cpc-hepnp}

\usepackage{multicol}
\usepackage{graphicx}
\usepackage{booktabs}
\usepackage{amssymb,bm,mathrsfs,bbm,amscd}
\usepackage[tbtags]{amsmath}
\usepackage{lastpage}

\begin{document}

\fancyhead[c]{Submitted to ``Chinese Physics C"} \fancyfoot[C]{\small \thepage}

\footnotetext[0]{Submitted to ``Chinese Physics C"}

\title{Detection efficiency evaluation for a large area neutron sensitive microchannel plate detector\thanks{Supported by National Natural Science
Foundation of China (11175098) }}

\author{%
Yi-Ming Wang $^{1,2}$
\quad Yi-Gang Yang $^{1,2;1)}$ \email{yangyigang@mail.tsinghua.edu.cn}%
\quad Ren Liu $^{1,2}$
}
\maketitle

\address{%
$^1$ Department of Engineering Physics, Tsinghua University, Beijing, 100084, P.R.China \\
$^2$ Key Laboratory of Particle $\&$ Radiation Imaging (Tsinghua University), Ministry of Education, Beijing, P.R.China\\
}

\begin{abstract}
In this paper, the detection efficiency of a large area neutron sensitive microchannel plate detector has been evaluated. A $^6$LiF/ZnS detector was employed as the benchmark detector, the TOF spectra of these two detectors were simultaneously measured and the energy spectra were then deduced to calculate the detection efficiency curve of the $^n$MCP detector. Tests show the detection efficiency@25.3 meV thermal neutron is 34$\%$ for this $^n$MCP detector.
\end{abstract}

\begin{keyword}
microchannel plate; neutron detector; detection efficiency
\end{keyword}

\begin{pacs}
PACS:28.20.Pr
\end{pacs}

\footnotetext[0]{\hspace*{-3mm}\raisebox{0.3ex}{$\scriptstyle\copyright$}2013
Chinese Physical Society and the Institute of High Energy Physics
of the Chinese Academy of Sciences and the Institute
of Modern Physics of the Chinese Academy of Sciences and IOP Publishing Ltd}%

\begin{multicols}{2}

\section{Introduction}
Nondestructive Testing(NDT) is of great importance to homeland security and national defense industry. As important NDT techniques, X-ray radiography and Gamma-ray radiography have been widely applied in many areas. Compared to X-ray radiography and Gamma-ray radiography, neutron radiography has a variety of advantages. Neutrons are much more sensitive to lighter elements, such as hydrogen, carbon and their compounds because neutrons have much larger attenuation coefficients for lighter elements than photons. Neutrons are able to penetrate high-Z materials, which are opaque to X-rays. Moreover, thermal neutrons are very sensitive to a few special nuclides, such as $^{3}$He, $^{6}$Li, $^{10}$B, $^{113}$Cd, $^{155}$Gd, and $^{157}$Gd. Therefore, neutron radiography is a very important NDT technique and has been used in various areas. For example, neutron radiography started to be applied in the detection of hidden shot balls in a gas-cooled turbine blade\cite{lab1} when the use of X-ray radiography seemed difficult.

\textbf{N}eutron-sensitive \textbf{M}icro \textbf{C}hannel \textbf{P}late ($^n$MCP) has played a very important role in neutron imaging due to its high spatial resolution. $^n$MCP detectors was proposed by G. W. Fraser et al.\cite{lab2} and have been used by various researchers to achieve high-quality neutron imaging.\cite{lab3}\cite{lab4}\cite{lab5}\cite{lab6}\cite{lab7} To expand our skill in neutron imaging, a $^n$MCP detector has been developed. To make the MCP neutron sensitive, the normal MCP with the diameter of 106 mm was doped with 3 mol\% of $^{nat}$Gd$_2$O$_3$.\cite{lab8} The delay line anode was used as the readout anode for its large area, high throughput and good spatial resolution.

Spatial resolution and detection efficiency are two crucial parameters to achieve high-quality neutron imaging for neutron sensitive microchannel plate detectors . The spatial resolution directly determines the quality of the neutron image. The detection efficiency, along with the intensity of the neutron beam, determines the image acquisition time for a neutron imaging detector. To acquire a high-quality neutron image with the microchannel plate detector, enough neutron counts are essential, adequate statistics of at least $\sim$100 neutrons per pixel are required.\cite{lab9}  Typical neutron flux of a neutron beam is $10^{6}-10^{7}n/cm^{2}\cdot s$, \cite{lab10}\cite{lab11} the pixel of the microchannel plate detector is $\sim$10 $\mu m$, the detector's diameter is 106 mm, and the microchannel plate's open area ratio is 56$\%$. So the image acquisition time is estimated to be 7 s to 70 s, with a fully ideal detection efficiency of 100$\%$. However, if the detection efficiency is degraded to 10$\%$, the image acquisition time would be as long as 70 s to 700 s, resulting in a time-consuming imaging process. Hence, it is important to evaluate and improve the detection efficiency to shorten the image acquisition time, in other words, to acquire more effective neutron counts with a given measurement time and to ultimately enhance the quality of the neutron image.

In our previous work, X-ray and Neutron images of an USAF-1951 Gd-mask have been acquired with this detector, the spatial resolution has been analysed and the throughput has been evaluated. This detector¡¯s spatial resolution is 62 $\mu m$ for 14 kV X-ray photons and is 88 $\mu m$ for thermal neutrons respectively\cite{lab12} In this paper, we present the principle, process and results of the thermal neutron detection efficiency measurement of this detector.

\section{The $^n$MCP detector}

The structure of the $^n$MCP detector is shown in figure 1. This detector is composed of a Gd-Mask, the MCP V-stack, the delay-line anode and its related electronics. The imaging object (Gd-mask) was designed as the USAF-1951 spatial resolution test chart. The MCP V-stack consists of a neutron sensitive MCP and a normal MCP. The delay line anode is a two dimensional readout anode with 120 mm by 120 mm area. The 4-channel readout electronics is composed of the differential amplifier, the CFD (constant fraction discriminator) and the TDC (time to digital convertor).

\begin{center}
\includegraphics[width=8cm]{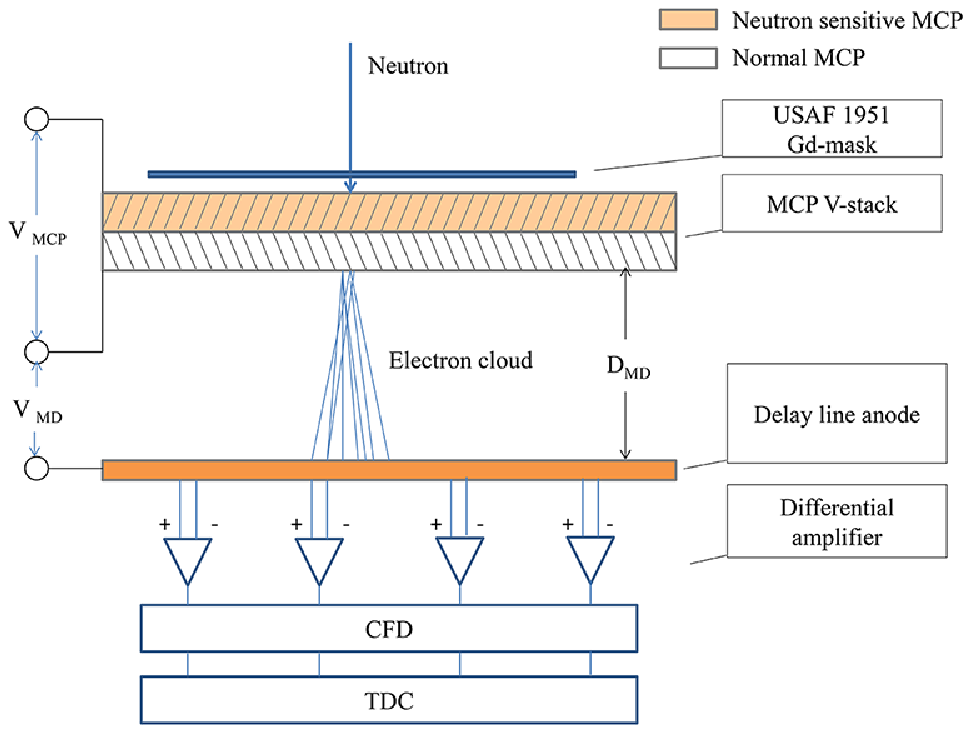}
\figcaption{\label{fig1} the structure of the neutron sensitive MCP detector with delay line readout. }
\end{center}

\section{The study of the thermal neutron detection efficiency}

\subsection{The principle of the detection efficiency measurement}
The detection efficiency measurements were undertaken at the thermal neutron beamline at the CPHS (\textbf{C}ompact \textbf{P}ulsed \textbf{H}adron \textbf{S}ource) at Tsinghua University. Considering that the neutron flux of the CPHS would fluctuated and cannot be monitored online precisely, a $^6$LiF/ZnS detector with known detection efficiency curve was employed as a benchmark detector. The $^6$LiF/ZnS detector and the $^n$MCP detector was adjacently placed at two beamlines of CPHS and the event counts of these two detectors were simultaneously measured. Figure 2 shows the two detectors layout.

\begin{center}
\includegraphics[width=8cm]{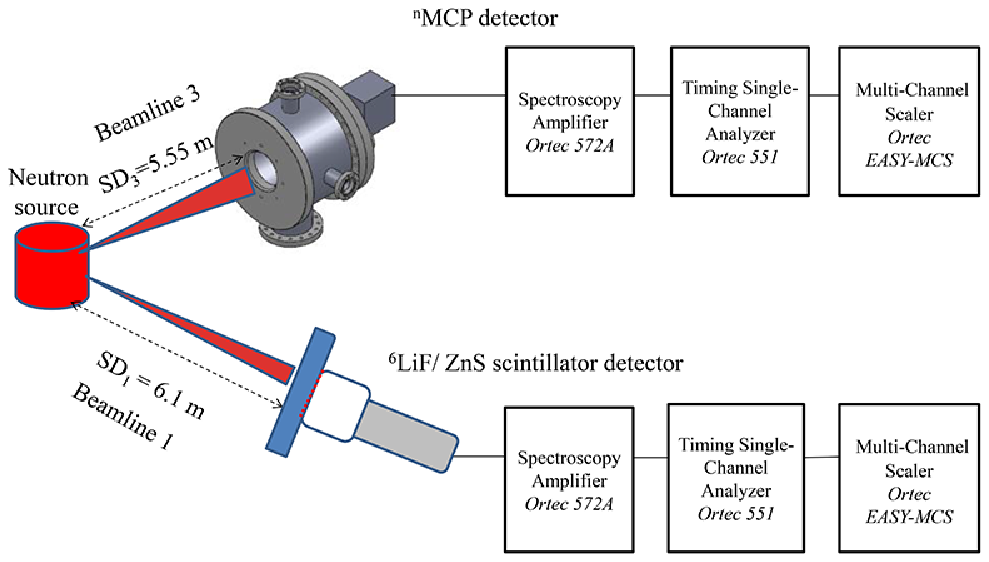}
\figcaption{\label{fig2}the $^6$LiF/ZnS detector and the $^n$MCP detector. }
\end{center}

The event counts ratio of the two detectors can be written as equation (1)

\begin{eqnarray}
\label{eq1}
\frac{n_{^nMCP}}{n_{^6Li}} = \frac{\epsilon_{^nMCP}}{\epsilon_{^6Li}}
{\left(\frac{SD_{^6Li}}{SD_{^nMCP}}\right)}^{2}
\frac{S_{^nMCP}}{S_{^6Li}}
\frac{1-\tau_{^6Li}\cdot n_{^6Li}}{1-\tau_{^nMCP}\cdot n_{^nMCP}}
\end{eqnarray}

where \emph{n} are the event counts, \emph{$\epsilon$} is the intrinsic detection efficiency of the detector, \emph{SD} is the distance between the neutron source and the detector, \emph{S} is the detector's effective detection area, and $\tau$ is the detector's dead time. In the measurement, $S_{^nMCP} = 33.8$ $cm^2$(subtract the Gd-mask's area from the detector's total area), $S_{^6Li} = 3.14$ $cm^2$, $SD_{^nMCP} = 5.55$ $m$, $SD_{^6Li} = 6.1$ $m$ and $\tau_{^nMCP} = \tau_{^6Li} = 3$ $\mu s$, which is determined by the amplifier's time constant for pulse-shaping. With equation (1), we have:

\begin{eqnarray}
\label{eq2}
{\epsilon_{^nMCP}}=0.0769 \cdot {\epsilon_{^6Li}} \cdot \frac{n_{^nMCP}}{n_{^6Li}} \cdot
\frac{1-\tau_{^nMCP}\cdot n_{^nMCP}}{1-\tau_{^6Li}\cdot n_{^6Li}}
\end{eqnarray}

Based on equation (2), the detection efficiencies at different neutron energies can be deduced by measuring the energy spectra of two detectors; hence, we measured the TOF(Time of Flight) spectra of these two detectors. The CPHS worked at pulsed mode and delivered neutrons with a 20 Hz frequency and a 100 $\mu$s duration time per period and the measurements were triggered by the synchronization signal of the accelerator.

Figure 3 shows the the TOF spectra of the $^n$MCP detector and the $^6$LiF/ZnS detector.
\begin{center}
\includegraphics[width=8cm]{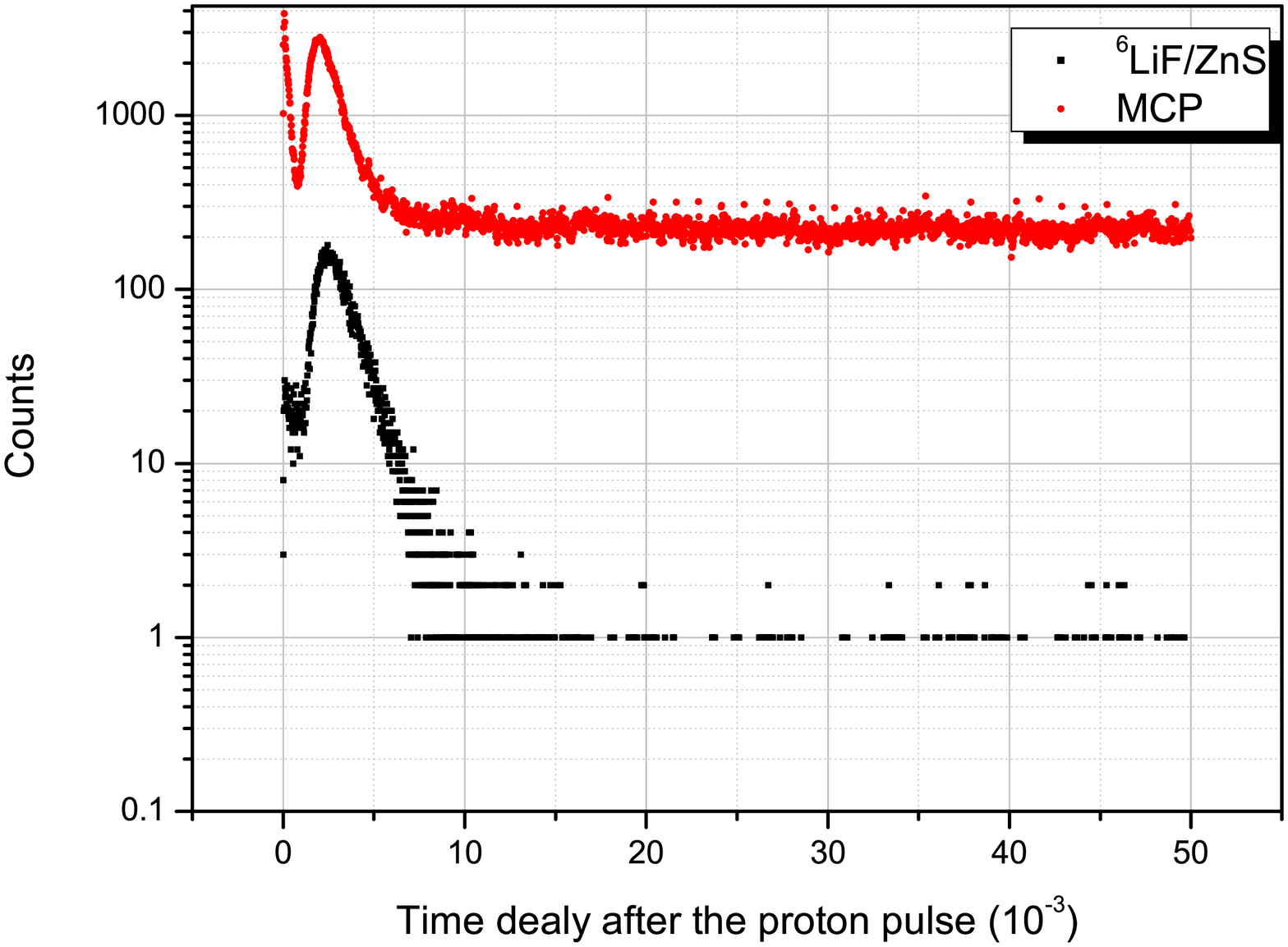}
\figcaption{\label{fig3}the TOF spectra of the $^6$LiF/ZnS detector and the $^n$MCP detector. }
\end{center}

As shown in figure 3, there were dark counts in the $^n$MCP detector TOF spectrum. To remove the dark counts, the counts with time of flight longer than 15 ms(neutron energy less than $0.71$ meV) were removed from the original spectrum. The TOF spectra were transformed as equation (3) to the energy spectra.
\begin{eqnarray}
\label{eq3}
E&=&\frac{1}{2}\cdot m \cdot {\left(\frac{SD}{t}\right)}^{2}
\end{eqnarray}
where \emph{E} is the neutron energy, \emph{m} is the neutron mass, \emph{SD} is the distance between the neutron source and the detector, \emph{t} is the time of flight. The energy spectra of these two detectors can be therefore acquired, as shown in figure 4.
\begin{center}
\includegraphics[width=8cm]{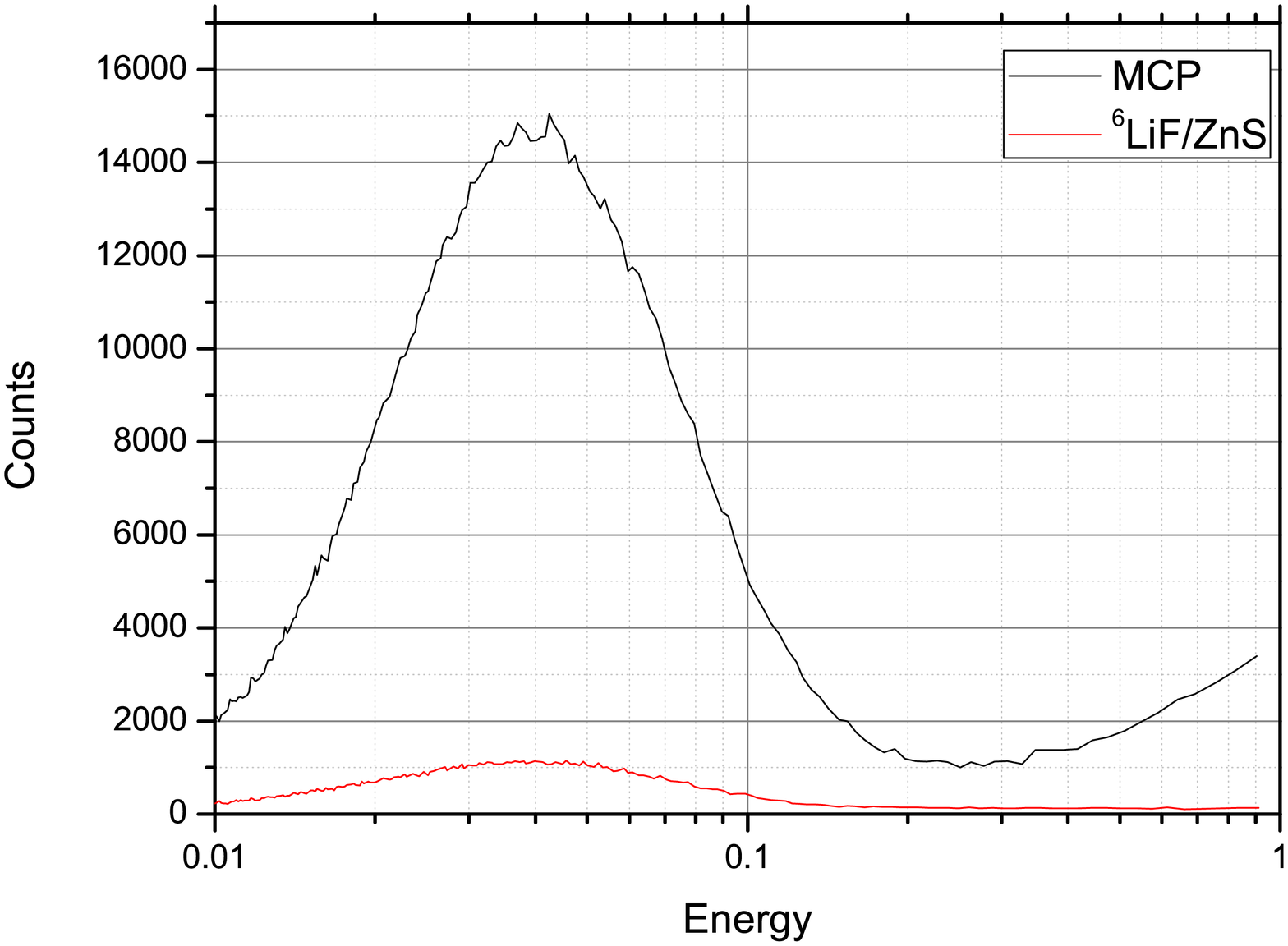}
\figcaption{\label{fig4}the energy spectra of the $^6$LiF/ZnS detector and the $^n$MCP detector. }
\end{center}

According to Equation (2), the detection efficiency curve of the $^n$MCP detector can be deduced from the event count ratios of the two detectors and the known detection efficiency curve of the $^6$LiF/ZnS detector. Figure 5 shows the calculated detection efficiency curve of the $^n$MCP detector and the know detection efficiency curve of the $^6$LiF/ZnS detector, the detection efficiency\emph{@}25.3\emph{meV }thermal neutrons is about 34$\%$.
\begin{center}
\includegraphics[width=8cm]{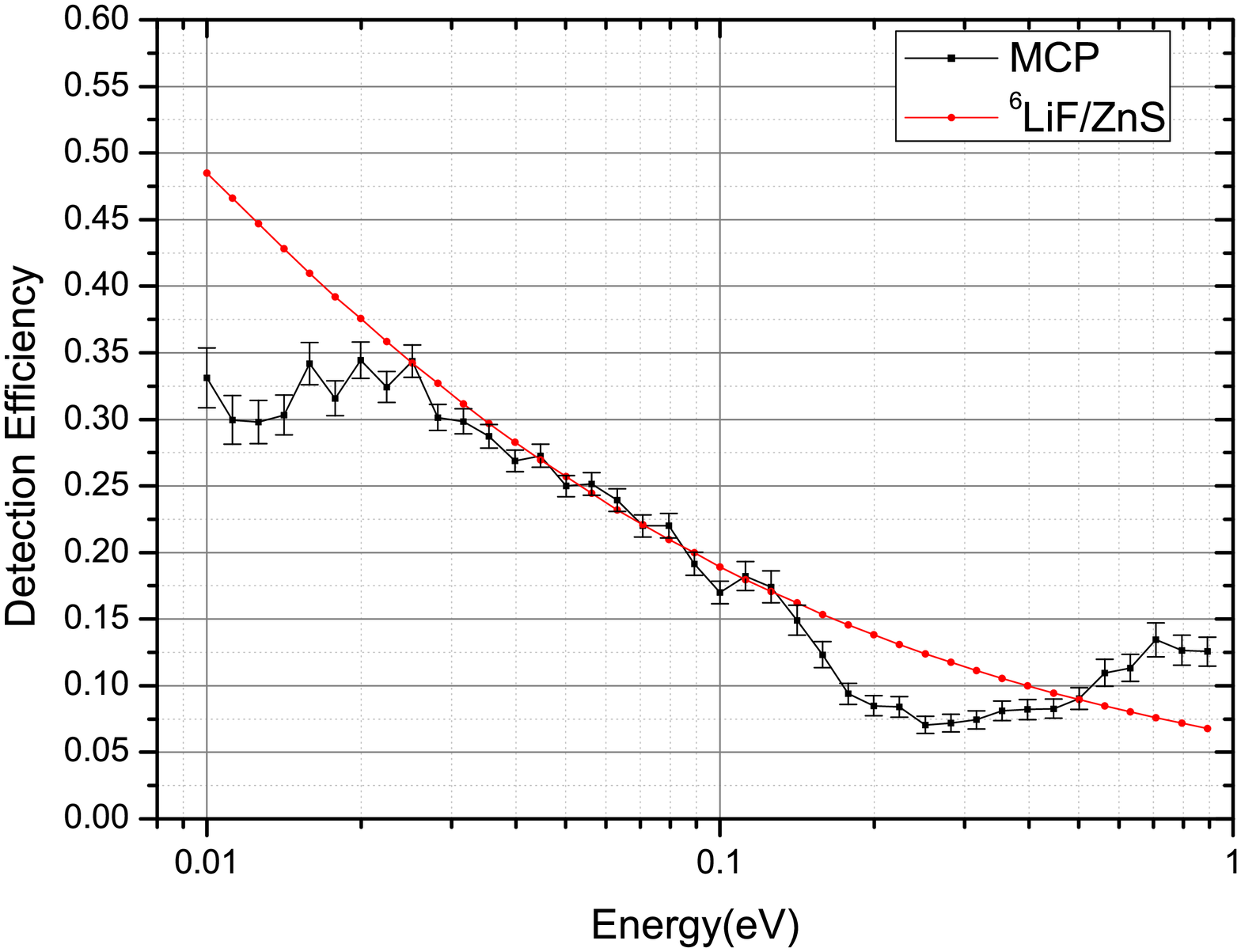}
\figcaption{\label{fig5}the detection efficiency curves of the $^6$LiF/ZnS detector and the $^n$MCP detector. }
\end{center}

\section{Conclusion and Discussion}

It is important to consider what effects has the 100 $\mu$s pulse width of the accelerator's synchronizing signal had on the $^n$MCP detector's detection efficiency measurements. This pulse width introduces a 100 $\mu$s error to TOF measurements, therefore introduces errors to neutron energy calculation. The relationship between energy error and TOF error is shown in equation (4).

\begin{eqnarray}
\label{eq4}
dE = m\cdot \left(\frac{SD^{2}}{t^{3}}\right)\cdot dt
\end{eqnarray}

Therefore, relative energy errors, $\frac{dE}{E}$, at different neutron energies can be deduced according to equation (4), as shown in figure 6.

\begin{center}
\includegraphics[width=8cm]{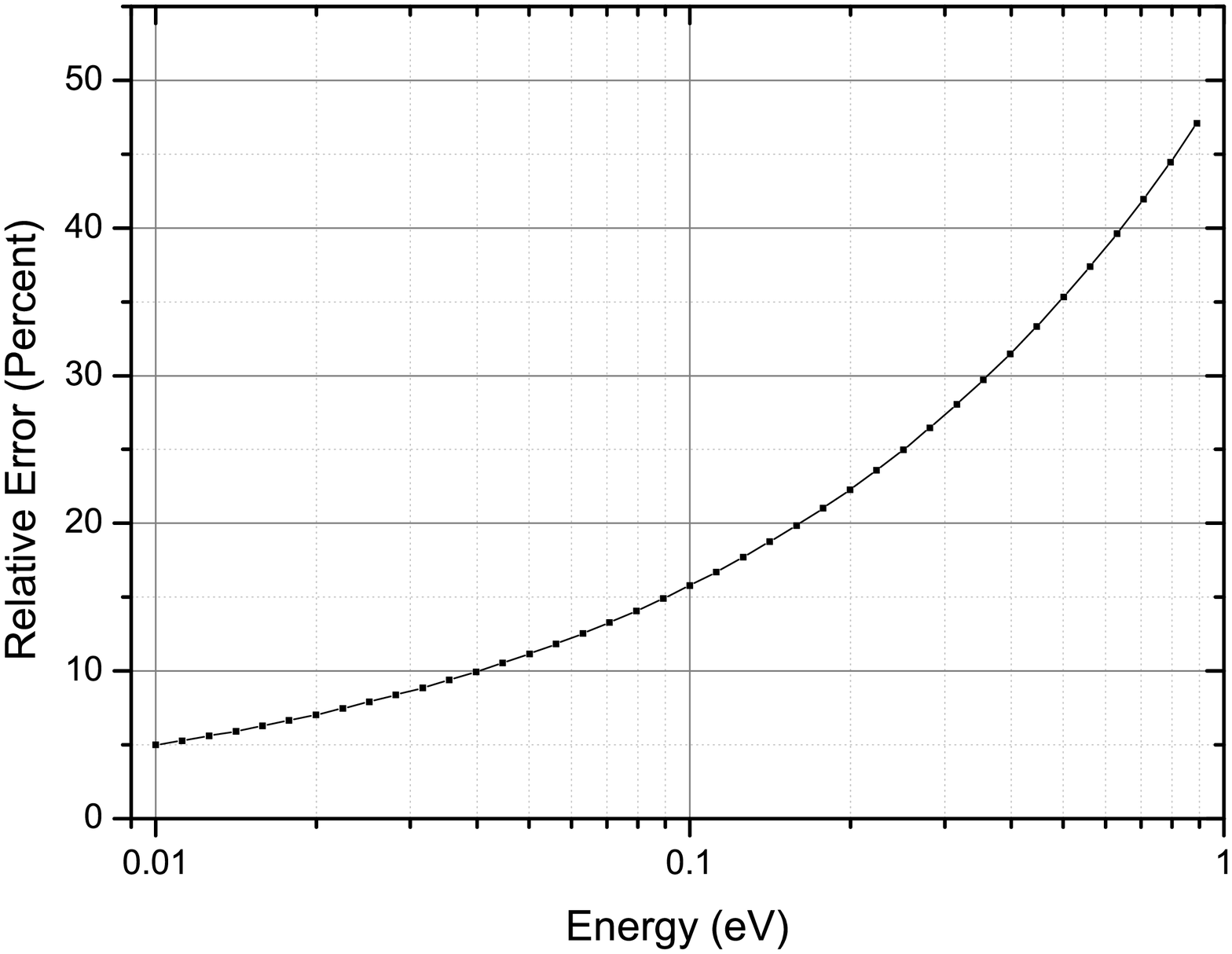}
\figcaption{\label{fig6} energy errors at different neutron energies. }
\end{center}

Taking neutron energy errors into consideration when drawing two detectors' detection efficiency curves, we have figure 7, which is the combination of figure 5 and figure 6.
\end{multicols}

\begin{center}
\includegraphics[width=18cm]{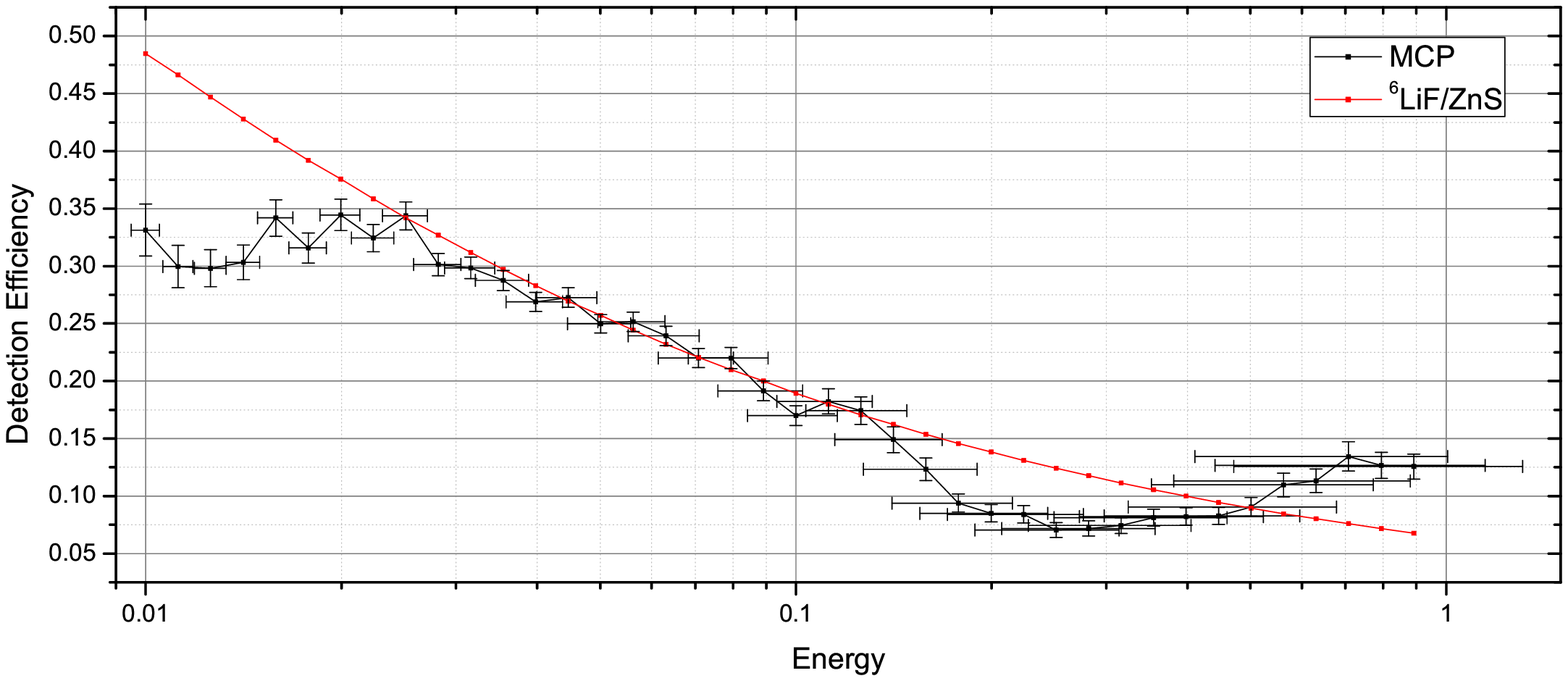}
\figcaption{\label{fig7} detection efficiency curves considering neutron energy error}
\end{center}

\begin{multicols}{2}
 Previous work on thermal neutron detection efficiency was conducted mainly by A.S. Tremsin et al. The thermal neutron detection efficiency of a $^{n}$MCP detector, which consisted of one $^{10}$B-doped microchannel plate followed by a stack of two normal MCPs, was measured to be 16$\%$, \cite{lab15} even though the calculation results predict that detection efficiencies of up to 78$\%$ are possible.\cite{lab13}\cite{lab14} Compared to previous detectors, our gadolinium doped microchannel plate detector has higher detection efficiencies for thermal neutrons, as our measurements results showed.

In conclusion, the detection efficiency of a large area neutron sensitive microchannel plate detector has been evaluated through measuring the TOF spectra of the detector and the benchmark detector, a $^6$LiF/ZnS detector with known detection efficiency curve. The energy spectra were deduced and the detection efficiency curve of the $^n$MCP detector were calculated. The detection efficiency\emph{@}25.3\emph{meV }thermal neutrons is about 34$\%$, which enable us to conduct neutron imaging experiments at low-intensity neutron beams like the CPHS.

\end{multicols}

\clearpage
\end{document}